# STUDY OF THE INFLUENCE OF INSTABILITY WAVE SCATTERING ON THE EFFICIENCY OF JET NOISE SHIELDING WITHIN THE GEOMETRIC THEORY OF DIFFRACTION

**O.P. Bychkov, S.A. Chernyshev, S.L. Denisov, G.A. Faranosov**

*Central Aerohydrodynamic Institute, Moscow, Russia*

The paper presents the results of a theoretical study of the noise shielding effect of a subsonic turbulent jet located near a rigid screen. The developed theoretical model, based on the methods of the Geometrical Theory of Diffraction (GTD), takes into account the effects of scattering of the hydrodynamic fluctuations of the jet near field at the edges of the screen and diffraction of acoustic perturbations generated by small-scale turbulence. Small-scale turbulent noise radiating pulsations are described within the framework of the correlation theory of jet noise in the form of compact quadrupole sources distributed along the jet axis. Hydrodynamic near-field fluctuations are approximated as a superposition of Kelvin-Helmholtz instability wave packets of different azimuthal numbers. Typical parameters of both types of fluctuations are adjusted using experimental data on the near and far fields of a free jet issuing from a profiled nozzle with a diameter of 0.04 m at a velocity of 181 m/s. The result of the jet noise shielding efficiency calculation, using the developed model, is in good agreement with the experimental data obtained for the jet-plate configuration, including correct reproduction of the low-frequency noise amplification (jet-plate installation noise) and the effect of high-frequency noise decrease (shielding itself).

*Keywords:* jet-wing interaction noise, Kelvin-Helmholtz instability waves, correlation theory of jet noise, geometric theory of diffraction (GTD)

# INTRODUCTION

Engine noise is the main factor influencing the overall aircraft community noise levels. For turbojet engines, one of the main sources of aircraft noise is a turbulent jet. Its location near the aircraft wing can lead both to noise reduction (shielding effect) and noise increase (interaction noise) [1-3].

Studying the effect of jet noise shielding, it is important to choose a model for noise generation by an isolated turbulent jet. Despite the long history of turbulent jet research, there is still no commonly accepted model for the noise generation process (e.g., [4]). However, for subsonic jets, as shown in [4], a model of random quadrupole sources possessing certain correlation properties (the so-called correlation model) has some advantages in terms of prediction of jet noise characteristics including such subtle features as its azimuthal content [5-7]. In [8], the correlation model made it possible to predict sound pressure levels of a jet shielded by a rectangular screen, which were qualitatively and, in some cases, quantitatively, consistent with experimental measurements.

The implementation of the geometric theory of diffraction (GTD) as a calculation tool for studying the shielding efficiency is due to several factors: firstly, as shown in [9-11], it is GTD that provides a correct description of the phase structure of the sound field near the shielding surfaces, which is especially important for acoustically non-compact noise sources. And secondly, GTD has been successfully validated on problems of diffraction of sound radiated by both point sources [12] and acoustically non-compact sources of aerodynamic nature [5, 13, 14]. Taken together, these factors allow the adaptation of GTD methods to solve the problem of shielding taking into account the scattering of instability waves on the elements of the airframe structure (especially on sharp edges) since it is this effect that plays the main role in the formation of the diffracted sound field in the geometric shadow zone and is responsible for the shielding efficiency.

The methods developed in [8, 13, 14] made it possible to use the concept of point acoustic sources, with the help of which the non-compact aerodynamic noise sources of interest are formed. The perturbation field generated by these sources can be described either by a set of monopole and dipole sources on the Kirchhoff surface [14] or by a set of quadrupole sources on the jet axis [5]. For such sources, the calculation of the shielding efficiency by flat polygonal rigid screens is carried out using the methods developed in [13]. This approach allows to calculate not only the shielding efficiency, but also to identify the distribution of the full sound field at various observation points, which makes it possible to determine the optimal position of the power plant from the point of view of aircraft community noise reduction.

As a development of [5], in the present work, in addition to the effect of jet noise shielding, the effect of low-frequency noise amplification associated with the appearance of the additional source that manifests itself symmetrically relative to the wing surface will also be considered [15-17]. This effect is explained by the scattering of the jet near-field hydrodynamic fluctuations on the wing trailing edge (a review of this effect is given in [18]). The development of the azimuthal decomposition technique, which allowed to conduct a thorough comparison of the azimuthal modes measured in the far field of an isolated high-speed jet with the predictions of the instability wave theory [19], made it also possible to obtain qualitative and quantitative [20-23] agreement between the jet-wing interaction noise model and experimental data on its directivity, spectrum and sensitivity to the relative position of the jet and the wing.

The problem of jet noise shielding assessment by the wing surface, considering the effect of low-frequency noise amplification, was considered in [20, 21]. However, in [20], the jet noise and hydrodynamic fluctuations were modeled in the framework of the unified wave-packet approach, which did not allow obtaining quantitative agreement with the measurements. In [21], acoustic and hydrodynamic fluctuations modeled separately, and satisfactory coincidence with the experiments was obtained. However, both the models for the fluctuations and the method for the scattering calculation were rather simplified. Jet noise sources were modeled without taking into account the detailed azimuthal structure of the real jet, and near-field perturbations were represented in the uniform approximation neglecting wave-packet envelope. The scattering problem in [21] was solved using a half-plane Green’s function in the form presented in [31], while the scattering of the hydrodynamic near field was considered using Amiet’s approach.

In the present study, we propose an approach combining detailed modeling of the fluctuation field incident on the screen (wave packets and azimuthally adjusted quadrupole sources) and quite precise solution of their scattering on the screen using GTD.

## MODELING OF A FREE JET ACOUSTIC AND HYDRODYNAMIC FIELDS

In general, subsonic jet noise can be described as a superposition of two main components [4, 30]:

1. Noise radiated by small-scale turbulence;
2. Noise radiated by large-scale coherent structures.

Both noise sources are broadband, but while the scattering of small-scale turbulence noise at the edge of the shielding surface has little effect on the total acoustic field, the scattering of the non-radiating part of the field of large-scale coherent structures begins to play a significant role as the jet approaches the screen. Even though these sources have different natures, they allow a uniform description in application to the problem of jet noise shielding.

A simplified correlation model that does not consider the effect of refraction of sound waves on the mean jet flow [5-7] is used. The detailed description of the used model of jet mixing noise can be found in [8]. Within the framework of this approach, acoustic perturbations are described using a simple wave operator based on the Lighthill acoustic analogy [24]:

$$\nabla^2 p - \frac{1}{c_0^2}\frac{\partial^2 p}{\partial t^2} = \frac{\partial^2 T^{ij}}{\partial r^i \partial r^j}, \qquad (1)$$

where $p$ is the pressure pulsations, $c_0$ is the speed of sound, $T^{ij}$ is a non-stationary part of the Reynolds stress tensor $\rho v^i v^j$. The right-hand side is described as a superposition of basic quadrupoles:

$$T^{ij}(\mathbf{r},t) = \sum_{n=1} D_n^{ij}\xi_n(\mathbf{r},\mathrm{t}), \qquad (2)$$

where $\xi_n$ are the stationary random fields with given correlation characteristics, and the basic quadrupoles $D_n^{ij}$ are selected such that $\xi_n$ for different $n$ are statistically independent [5-7]. In this case, the sound pressure spectrum can be represented as:

$$\Phi_p(\mathbf{r},\omega) = \sum_{n=1}^{6}\int D_n^{ij} G^{ij}(\mathbf{r},\mathbf{r}',\omega) D_n^{kl} G^{kl*}(\mathbf{r},\mathbf{r}'',\omega)\Phi_{\xi n}(\mathbf{r}',\mathbf{r}'',\omega)\,\mathrm{d}\mathbf{r}'\mathrm{d}\mathbf{r}'', \qquad (3)$$

where $G^{ij}(\mathbf{r},\mathbf{r'},\omega)$ is the second derivative of the Green's function for the boundary value problem under consideration, calculated from the coordinates of the observation point, $\Phi_{\xi n}(\mathbf{r'},\mathbf{r''},\omega)$ is the cross spectral density function of a random field $\xi_n$, superscript * denotes a complex conjugation, and the integration is carried out over the volume occupied by the jet.

The common approach consists in the transition, within the framework of the Langevin equation, to the description of the characteristics of the sound sources using the intensity of fluctuation generation. The choice of the correlation function of perturbation generation $\eta(\mathbf{r},t)$ in the form of:

$$R_\eta(\mathbf{r}',\mathbf{r}'',\tau) = A(\mathrm{r}')\exp\left(-\frac{|r'-r''|^2}{2l_0^2} - \frac{\tau^2}{2\sigma^2}\right), \qquad (4)$$

where $l_0$ is the correlation length and is considered small in comparison with other characteristic dimensions of the problem, and the projection of the noise sources on the jet axis, allow to obtain the following relationship for the sound pressure spectrum (3):

$$\Phi_p(\mathbf{r},\omega) = \sum_{n=1}^{6}\int D_n^{ij} G^{ij}(\mathbf{r},\mathbf{r}_2,\omega) D_n^{kl} G^{kl*}(\mathbf{r},\mathbf{r}_1,\omega)\hat{H}(x_2,x'',\omega)\hat{H}^*(x_1,x',\omega)\Phi_\eta(x',x'',\omega)dx'dx''dx_1dx_2, \qquad (5)$$

where $\hat{H}(x,x_0,\omega)$ is the Fourier image of the Green's function of the Langevin equation, which has the form:

$$\hat{H}(x, x_0, \omega) = \int \exp(-i\omega\tau) H(x, x_0, \tau) d\tau,$$
$$H(x, \tau) = \frac{1}{U} \delta(x - U\tau) \exp(-\varepsilon\tau) \theta(\tau), \qquad (6)$$

where $U$ is a convective speed of sources, $x$ is a longitudinal coordinate along the jet axis, $\varepsilon(x)$ is a perturbation attenuation rate. The cross-spectral function $\Phi_\eta$ in (5) is written as follows:

$$\Phi_\eta(\mathbf{r}', \mathbf{r}'', \omega) = 4\pi^2 A(\mathbf{r}') \sigma l_o^3 \delta(\mathbf{r}' - \mathbf{r}'') \exp\left(-\frac{\omega^2 \sigma^2}{2}\right), \qquad (7)$$

where $\delta(\mathbf{r})$ is the Dirac delta function and $\theta(t)$ is the Heaviside function.

The main advantage of expression (5) is its explicit dependence on the Green's function, which allows it to be used not only to calculate the propagation of jet noise in free space, but also in the presence of various reflective and/or shielding surfaces (the so-called installed jet). Considering Green’s function for the case of free space, it is possible to select the necessary parameters used in the model based on experimental data on far-field noise.

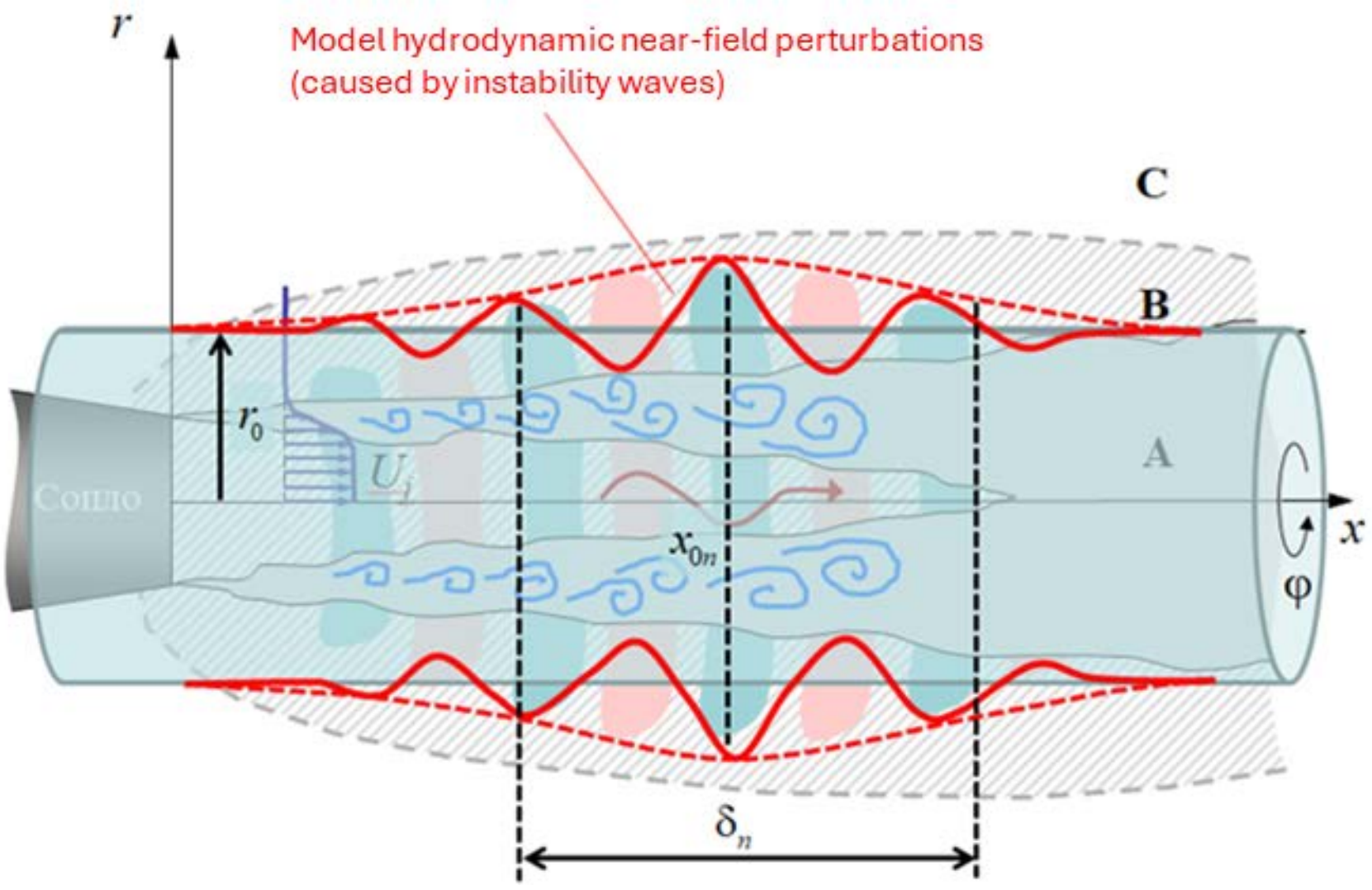

Fig. 1. On the modeling of large-scale structures in the jet near field

Modeling of hydrodynamic fluctuations of the jet near field is performed in accordance with the approach described in [22]. Unlike the field of small-scale sources, represented as a superposition of point quadrupoles, the pressure field from large-scale instability waves is specified as a solution to a boundary value problem for each frequency ω and each azimuthal mode $m$, in which the distribution of the perturbation intensity is considered to be known on some surface located in the vicinity of the jet, for example, on a cylinder of radius $r_0$, surrounding the jet (Figure 1):

$$p_i(x, r, \varphi)\Big|_{r=r_0} = \sum_{m=-1}^{1} A'_m e^{-\frac{(x-x_m)^2}{\delta_m^2} + i\omega\frac{x-x_m}{V_m} + im\varphi}, \qquad (8)$$

where $\delta_m$ is a wave packet width (Gaussian type), $x_m$ is a position of the maximum of the packet, $V_m$ is a convective speed of perturbations, $A'_m$ is an amplitude of the pulsations at a given frequency for the azimuthal mode $m$. Following the results of works [21, 22], the summation in (8) is carried out only for the axisymmetric and first azimuthal harmonics, since they make the main contribution to the near-field pressure fluctuations for typical relative jet-wing positions.

Outside the cylindrical surface, the pressure perturbation satisfies the Helmholtz equation:

$$\Delta p + k^2 p = 0, \qquad r > r_0, \tag{9}$$

with boundary condition (8) and the radiation condition at infinity, where $k = \omega/c$ is a wave number. The final expression for the determined perturbations can be written as follows:

$$p_i(x,r,\varphi) = \sum_{m=-1}^{1} e^{im\varphi}\delta_m \frac{A_m}{2\sqrt{\pi}} \int_{-\infty}^{\infty} \frac{K_m(\gamma r)}{K_m(\gamma r_0)} e^{-i\alpha(x-x_m)-\frac{\delta_m^2}{4M_m^2}(k+M_m\alpha)^2} d\alpha, \qquad \gamma = \sqrt{\alpha^2 - k^2}, \tag{10}$$

where $\alpha$ is a longitudinal wave number, $M_m = V_m/c$ is an acoustic Mach number, $c$ is a speed of sound, $K_m$ is a modified Bessel function of the second kind of the order $m$, square root $\gamma$ is determined by that $\gamma(0) = -ik$, and the branch cuts go down from the point-$k$ to infinity in the lower half-plane, and from the point $k$ up to infinity in the upper complex half-plane parallel to imaginary axis.

Following the results of work [23,24], a further simplification of the model was made, namely it was assumed that the parameters of the axisymmetric mode and the first azimuthal modes are close to each other. Then expression (10) takes the following form:

$$p_i(x,r,\varphi) \approx \frac{A_t\delta_0}{2\sqrt{\pi}} \int_{-\infty}^{\infty} \frac{K_0(\gamma r)}{K_0(\gamma r_0)} e^{-i\alpha(x-x_0)-\frac{\delta_0^2}{4M_0^2}(k+M_0\alpha)^2} d\alpha, \tag{11}$$

where $A_t$ is an amplitude of the total near-field fluctuations.

## DESCRIPTION OF THE METHOD FOR NOISE SHIELDING ESTIMATION AT GIVEN NON-COMPACT SOURCE

This paper uses a method for calculating the shielding of an aerodynamic noise source of an arbitrary type, which is based on the approaches implemented within the framework of the Geometric Theory of Diffraction [25,26].

This method was described in detail in [5] and below we present a description of its modification for the case of non-compact sources consideration. Since large-scale coherent structures of the wave packet type described by expression (11) are non-compact, it is natural to use the approach developed in [14], which uses Kirchhoff's integral theorem, for their correct description. Due to the geometry of the system under consideration, it is most convenient to choose the Kirchhoff surface in the form of a cylinder, coaxial with the jet and located in the vicinity of the jet outer boundary (Figure 2).

Strictly speaking, the length of the Kirchhoff surface $L$ should be infinitely large or end with cylindrical tips that include the nozzle and the jet. However, this paper considers a simplified (practically realizable) Kirchhoff surface extending over several dozens of wavelengths in the upstream and downstream directions from the nozzle exit. For each frequency under consideration, the length of the Kirchhoff surface is selected from the condition that the amplitude of the pressure pulsations and their first derivative at the edges of the surface are two orders of magnitude smaller than the maximum value. Such a condition, on the one hand, allows to avoid complications when constructing the Kirchhoff surface, and on the other hand, to significantly speed up and simplify the calculation of the noise generated by the interaction of the large-scale coherent structures with a scattering surface.

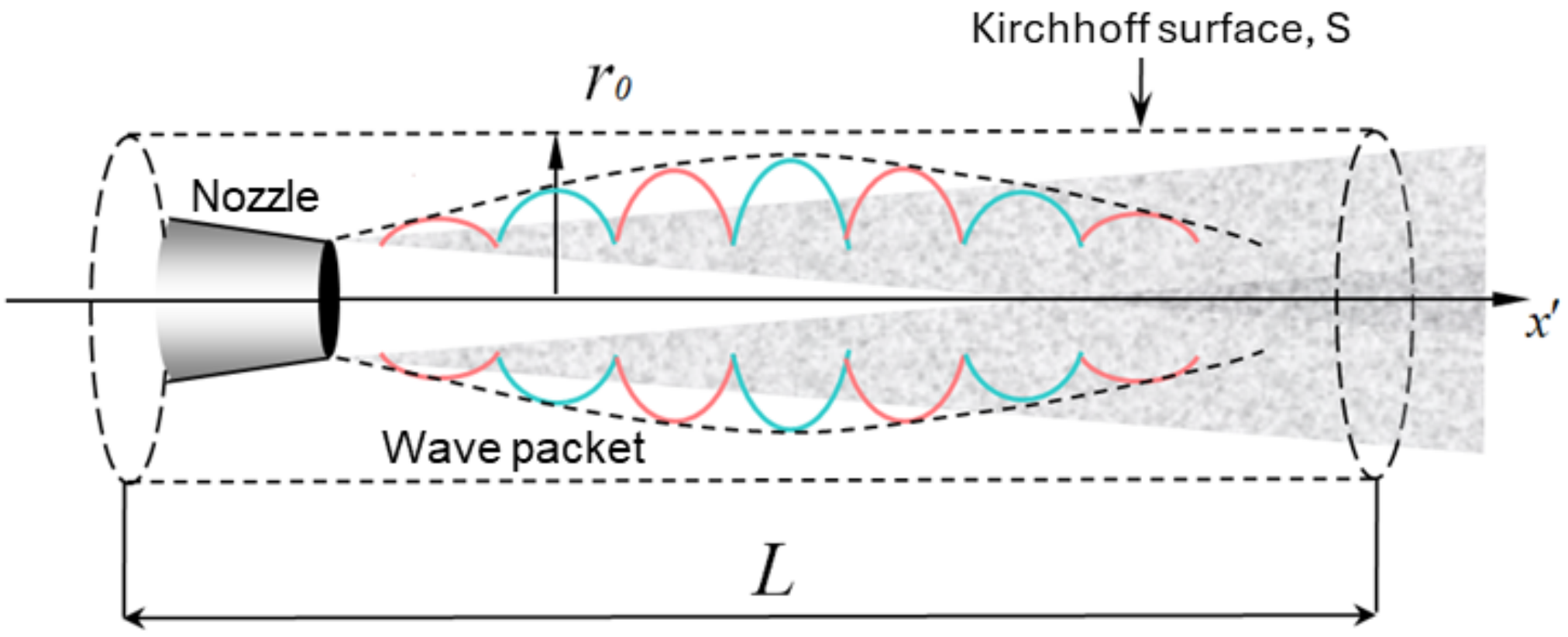

Fig. 2. On the construction of the Kirchhoff surface for calculating the noise shielding efficiency in case of large-scale coherent structures

The number of monopole-type point sources located on the Kirchhoff surface is determined from the condition of convergence of the results with an increase in their number. This number also depends on the frequency, the order of the considered azimuthal mode, and the width of the wave packet. For the convenience of calculations, the number of point sources was fixed for all frequencies and widths of wave packets.

To calculate the pressure fluctuation field $P(\mathbf{X}_{obs})$ for the large-scale coherent structures at an arbitrary observation point $\mathbf{X}_{obs}$, we apply Kirchhoff's integral theorem [14, 26] to the cylindrical surface enclosing the nozzle and the jet. As a result, we obtain the following expression:

$$P\left(\mathbf{X}_{obs}\right)=\int_{S}\left(G\left(\mathbf{X}_{obs},\mathbf{x}_{s}\right)\frac{\partial P}{\partial n}-P\frac{\partial G\left(\mathbf{X}_{obs},\mathbf{x}_{s}\right)}{\partial n}\right)d\sigma\,, \tag{12}$$

where the integration is performed over the entire cylindrical surface $S$ (the Kirchhoff surface), dσ is the surface element, the expressions for $P$ and $\partial P/\partial n$ on the surface $S$ describe the distribution of pressure pulsations and their first normal derivative. The functions $G$ and $\partial G/\partial n$ in equation (12) are the Green's function and its normal derivative to the Kirchhoff surface, which are determined from the solution of the Helmholtz equation for the boundary value problem under consideration in the presence of a shielding surface. Determining the functions $P$, $G$ and their normal derivatives in expression (12), one can calculate the pressure pulsations field for large-scale coherent structures at an arbitrary observation point in the presence of a shielding surface.

The distribution of pressure pulsations $P$ on a cylindrical surface for instability wave packets was obtained earlier in the form of expression (11). By differentiating this expression with respect to radius $r$, we can obtain the expression for $\partial P/\partial n$:

$$\frac{\partial P(x,r,\varphi)}{\partial r}=-\frac{A_{t}\delta_{0}}{2\sqrt{\pi}}\int_{-\infty}^{\infty}\gamma\frac{K_{1}\left(\gamma r\right)}{K_{0}\left(\gamma r_{0}\right)}e^{-i\alpha\left(x-x_{0}\right)-\frac{\delta_{0}^{2}}{4M_{0}^{2}}\left(k+M_{0}\alpha\right)^{2}}d\alpha\,. \tag{13}$$

Let us further consider the definition of the functions $G$ and $\partial G/\partial n$ included in expression (12) in the presence of a shielding surface. As is known from [25,26], in the presence of an acoustically rigid screen, the full sound field $G$ of a monopole source at an arbitrary observation point (Green's function) is a solution of the Helmholtz equation (9) with a non-zero right-hand side and is represented in the following form:

$$G=P_{G}+P_{D}, \tag{14}$$

where $P_G$ is a geometric-acoustic field, and $P_D$ is a diffracted field. For the geometric-acoustic field of the monopole source, the following expression is valid:

$$P_{G}=A\frac{\exp\left(ikR_{obs}\right)}{4\pi R_{obs}}\theta(R_{obs}),$$
$$R_{obs}=\sqrt{(X_{obs}-x_{s})^{2}+(Y_{obs}-y_{s})^{2}+(Z_{obs}-z_{s})^{2}}, \tag{15}$$

where $A$ is an initial field amplitude, $k$ is a wavenumber, $\boldsymbol{X}_{obs}$ represents an observation point coordinates, $\boldsymbol{x}_s$–the source coordinates (Figure 3), $R_{obs}$ is the radius vector, drawn from the source to the observation point, and $\theta(R_{obs})$ is a Heaviside step function.

For a monopole-type source considered within the framework of GTD, the expression for the diffracted field $P_D$ has the following form:

$$P_D = A\frac{\exp\left(ik\left(R_{sc}+R_d\right)\right)}{4\pi R_{sc}}\cdot\sqrt{\frac{R_{sc}}{R_d\left(R_{sc}+R_d\right)}}\cdot D_{coeff},$$
$$R_{sc} = \sqrt{(X_{sc}-x_s)^2+(Y_{sc}-y_s)^2+(Z_{sc}-z_s)^2}, \qquad (16)$$
$$R_d = \sqrt{(X_{obs}-X_{sc})^2+(Y_{obs}-Y_{sc})^2+(Z_{obs}-Z_{sc})^2},$$

where $\boldsymbol{X}_{sc}$ denotes coordinates of the scattering point on the edge of the screen, $R_{sc}$ is a radius vector of the scattering point drawn from the source to the scattering point on the edge of the screen, $R_d$ – radius vector of the observation point, $D_{coeff}$ – diffraction coefficient, determined from the solution of the canonical diffraction problem (more details are given in [8]). In the case under consideration, the diffraction coefficient is determined from the solution of the problem of diffraction of the sound field radiated by a monopole source on an acoustically rigid semi-infinite plane. Expressions (15), (16) completely describe the field of pressure pulsations at various observation points (in the lit zone, the shadow zone, and transition region).

Then, the expression for the derivative $\partial G/\partial n$ can be obtained by differentiating expressions (15) and (16) with respect to the source coordinates (this approach is described in more detail in [14]). Note that by performing repeated differentiation of the Green's function (14) with respect to the source coordinates, we obtain an expression that is used in relation (5) for calculating the spectrum of noise radiated by small-scale turbulence [8].

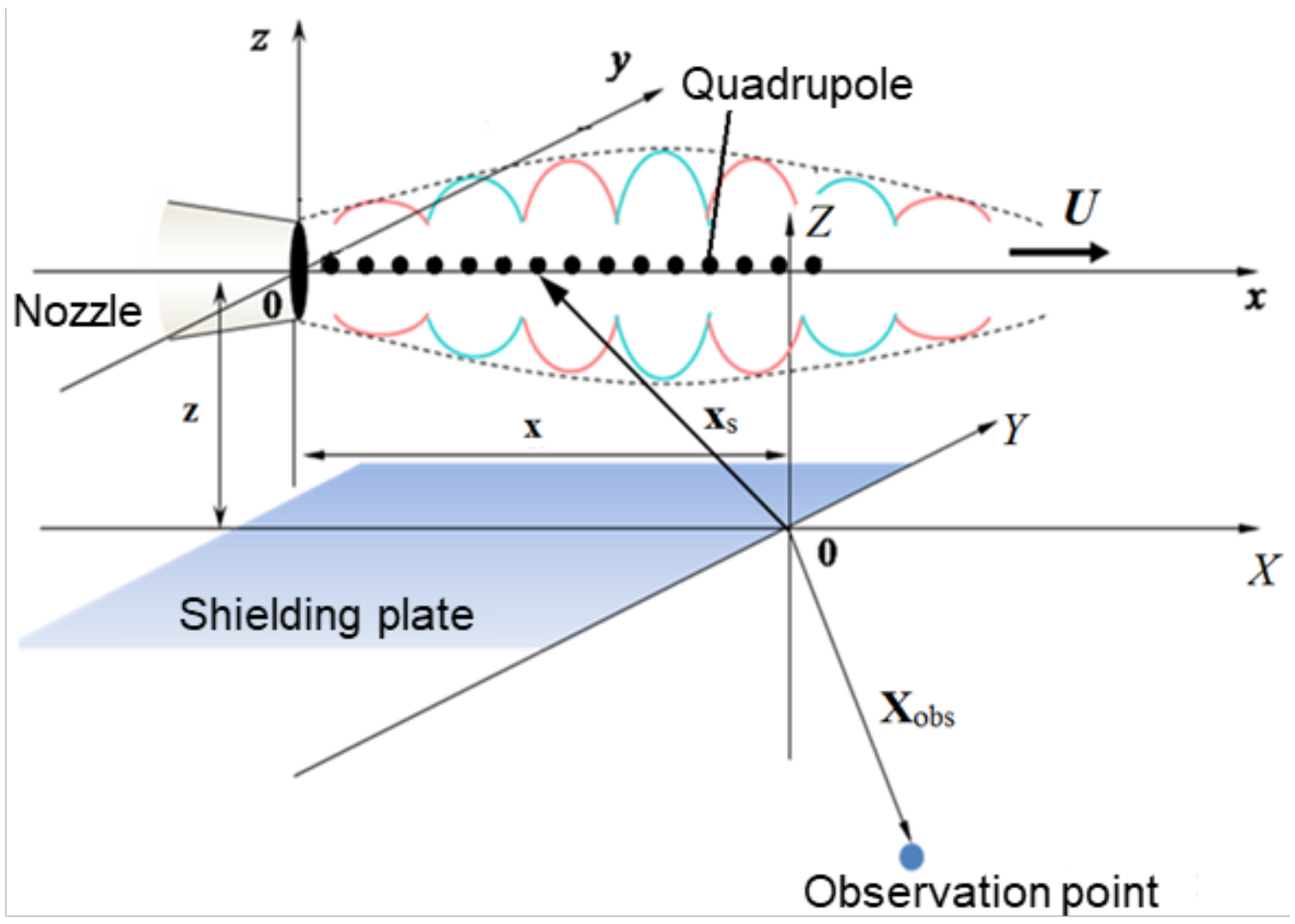

Fig. 3. Sketch for calculating the noise shielding of a round turbulent jet

To take into account the broadband nature pressure fluctuations from large-scale coherent structures, calculations using (12) are performed for a given frequency of the sound field ω, followed by squaring the calculated value of the mean sound field at this frequency and dividing it by the bandwidth Δω. This approach allows us to obtain the mean spectral density of the field in a given frequency band. By performing calculations for different frequencies with fixed or variable step Δω, one can determine the spectrum of large-scale structures self-noise and the noise from their interaction with the scattering surface in any desired frequency range, which can be compared with the experimental data.

## EXPERIMENTAL DATA FOR VALIDATION OF THE DEVELOPED METHOD

To validate the developed approach for calculating the shielding efficiency, a dataset from two series of experiments was used. A turbulent jet issuing at a velocity of $V_j = 181$ m/s from a nozzle with diameter $D = 0.04$ m was considered. To obtain empirical data on the characteristics of the coherent structures in the jet near field, the data of the experiment described in [22] were used (Figure 4a). The simultaneous use of a large number of microphones arranged in the form of rings made it possible to obtain the necessary information on the dominant azimuthal modes of instability wave packets in the region of interest in the near field of the jet.

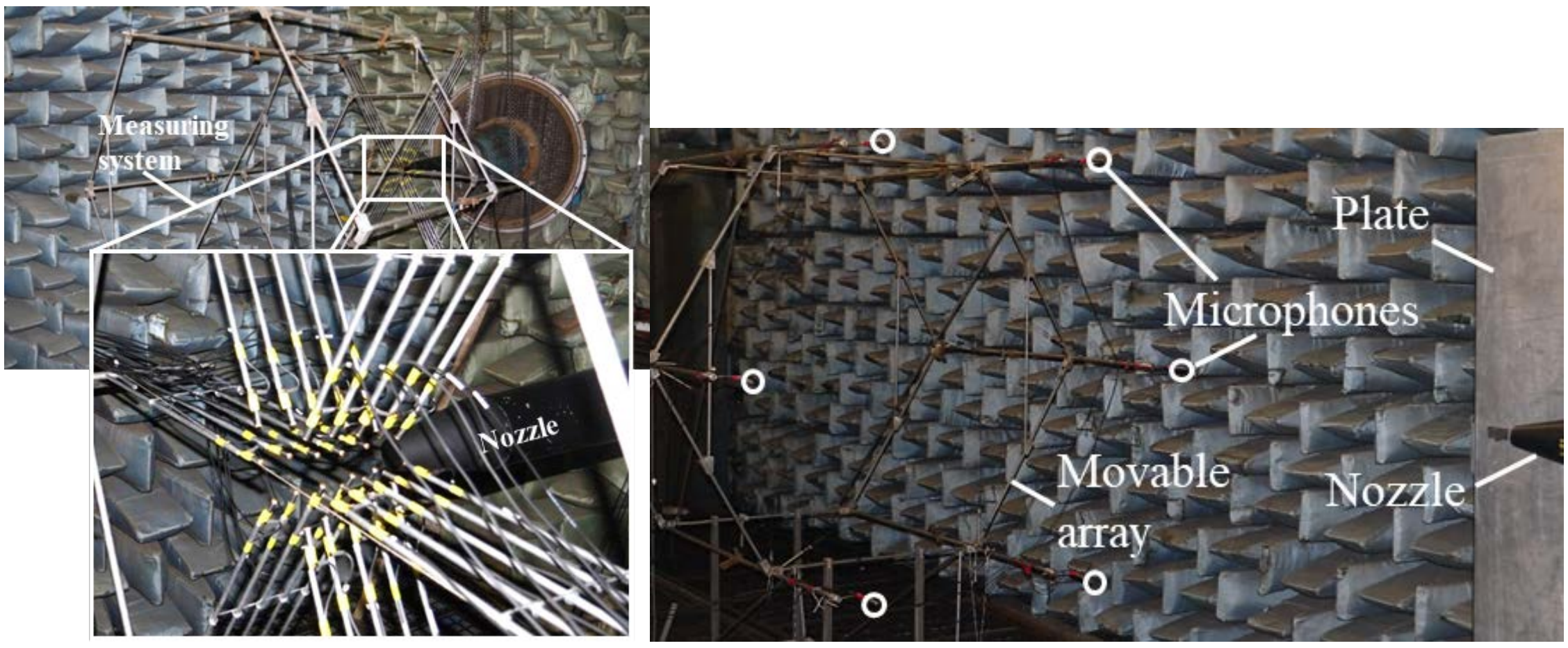

(a) (b)

Fig. 4. Experimental study in the AC-2 anechoic chamber. a) study of the near field of free jet; b) study of the far acoustic field of the jet and the jet with a rigid plate [22]

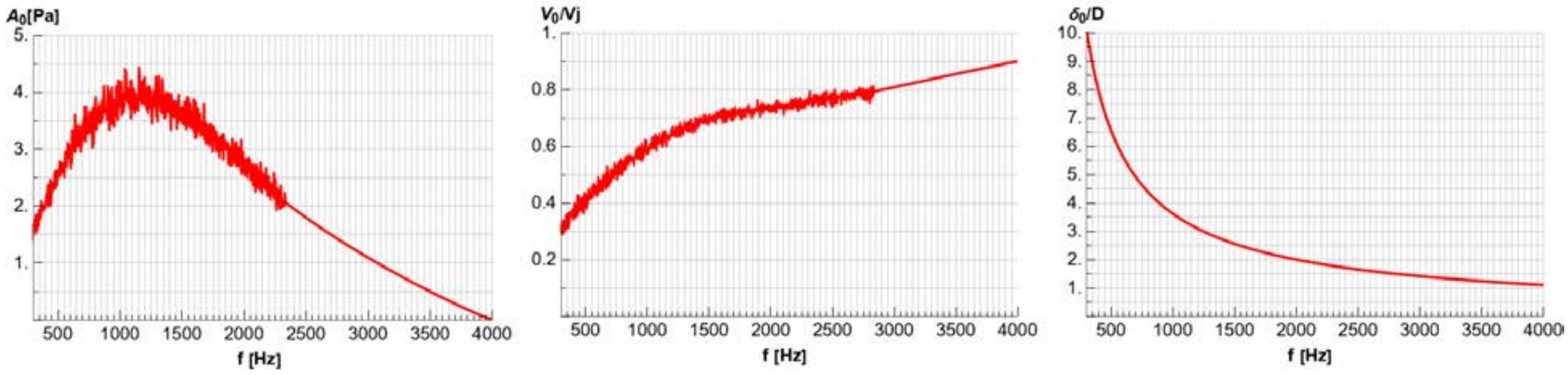

Fig. 5. Parameters of the large-scale structures in the jet near field for $r_0 = 0.9\,D$

Figure 5 shows the main characteristics of the wave packets determined in the analysis of the experimental data and used in the present calculations. The values of the amplitude and convective velocity for frequencies exceeding 2350 Hz (when the near-field hydrodynamic pulsations ceased to dominate the total field) were obtained using extrapolation, which ensures the constancy of the slope of the high-frequency part of the spectrum on a logarithmic scale. The position of the packet maximum was considered fixed and independent of the frequency $x_0 = 3\ D$, which can be due to the use of the Amiet approach [21, 27]. From the presented data, it can be expected that the scattering of instability wave packets will make the main contribution to the far acoustic field in the low-frequency region. The parameters necessary for the small-scale turbulence modeling in the form of a distribution of quadrupoles (5) were used from the previous work [8], in which an acceptable accuracy of the isolated jet noise modeling was demonstrated.

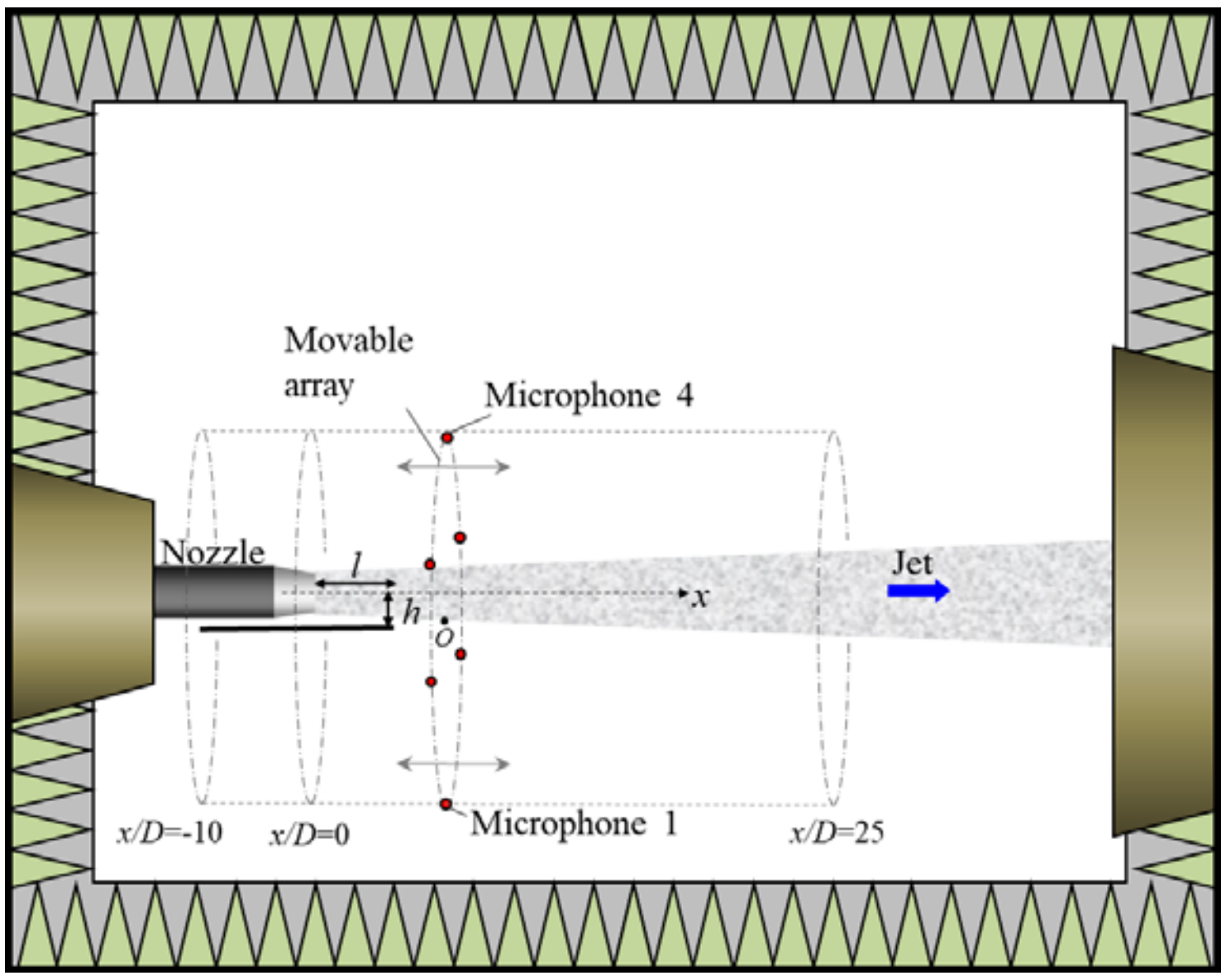

Fig. 6. Schematic representation of the far-field noise measurement area

It should be noted that the parameters of the axisymmetric mode could be determined using hot-wire measurements, which turns out to be much more accessible under conditions such as the presence of a concurrent flow [28].

The second series of experiments involved studying the far acoustic field of free and shielded jet [22] by means of a moving azimuthal microphone array (Figure 4b). The use of this measuring system made it possible to obtain information about the distribution of acoustic pressure pulsations on a cylindrical surface with a radius of $R = 0.8$ m surrounding the studied configuration, in the range of longitudinal coordinate values $-10 < x/D < 25$ (Figure 6). The distance from the nozzle exit to the trailing edge of the screen was described by the following parameter values: $l/D = 3$, $h/D = 1$ (details are presented in [29]).

# MODEL VALIDATION AND ANALYSIS OF THE JET NOISE SHIELDING EFFICIENCY CALCULATION

The algorithm for jet noise shielding efficiency calculation by airframe elements, taking into account the interaction of large-scale coherent structures (instability waves) with a screen, can be presented as a set of three main stages:

1. Calculation of free jet noise using correlation theory for small-scale turbulence noise and Kirchhoff's integral theorem for the self-noise of the instability waves;
2. Calculation of jet noise in the presence of a plane screen using Kirchhoff's integral theorem combined with GTD for both types of sources;
3. Calculation of shielding efficiency at observation points.

An important feature of both noise sources under consideration is the fact that they are stochastic and incoherent with respect to each other. This allows us to use energy summation in the considered frequency bands. Then the final expression for calculating the jet noise shielding efficiency $E_{sf}$ for the given frequency takes the form:

$$E_{sh} = 10\lg\left(\frac{\left(\Phi_{corr}\left(\mathbf{r},\omega\right)\big|_{screen} + P^2_{coherent}\left(\mathbf{r},\omega\right)\big|_{screen}\right)}{\left(\Phi_{corr}\left(\mathbf{r},\omega\right)\big|_{free} + P^2_{coherent}\left(\mathbf{r},\omega\right)\big|_{free}\right)}\right), \qquad (17)$$

where $\Phi_{corr}(\mathbf{r},\omega)|_{screen}$ and $\Phi_{corr}(\mathbf{r},\omega)|_{free}$ are the sound pressure spectra calculated using the correlation theory of jet noise (3) with and without the screen, respectively, and $P^2_{coherent}(\mathbf{r},\omega)|_{screen}$ with $P^2_{coherent}(\mathbf{r},\omega)|_{free}$ are the squared sound pressure levels from the coherent structures calculated using Kirchhoff's integral theorem (12) with and without ($P_D$=0) the screen, ($\mathbf{r}$,ω) are the coordinate of the observation point and the frequency of the sound field. It should be noted that $P^2_{coherent}(\mathbf{r},\omega)|_{free}$ turns out to be small due to the essentially subsonic convective velocity of the large-scale structures (for example, [4]). Calculations using the correlation theory of jet noise (3) and using the Kirchhoff integral theorem (12) are carried out in the same frequency bands $\Delta\omega$.

Figures 7-8 show the results of direct measurements and modeling using the developed algorithm of power spectral densities of jet noise in the far field on a cylindrical surface in several characteristic sections in the presence and absence of the shielding plate. The calculated values were obtained for a set of frequencies lying in the region from 300 Hz to 4000 Hz covering the spectral maximum.

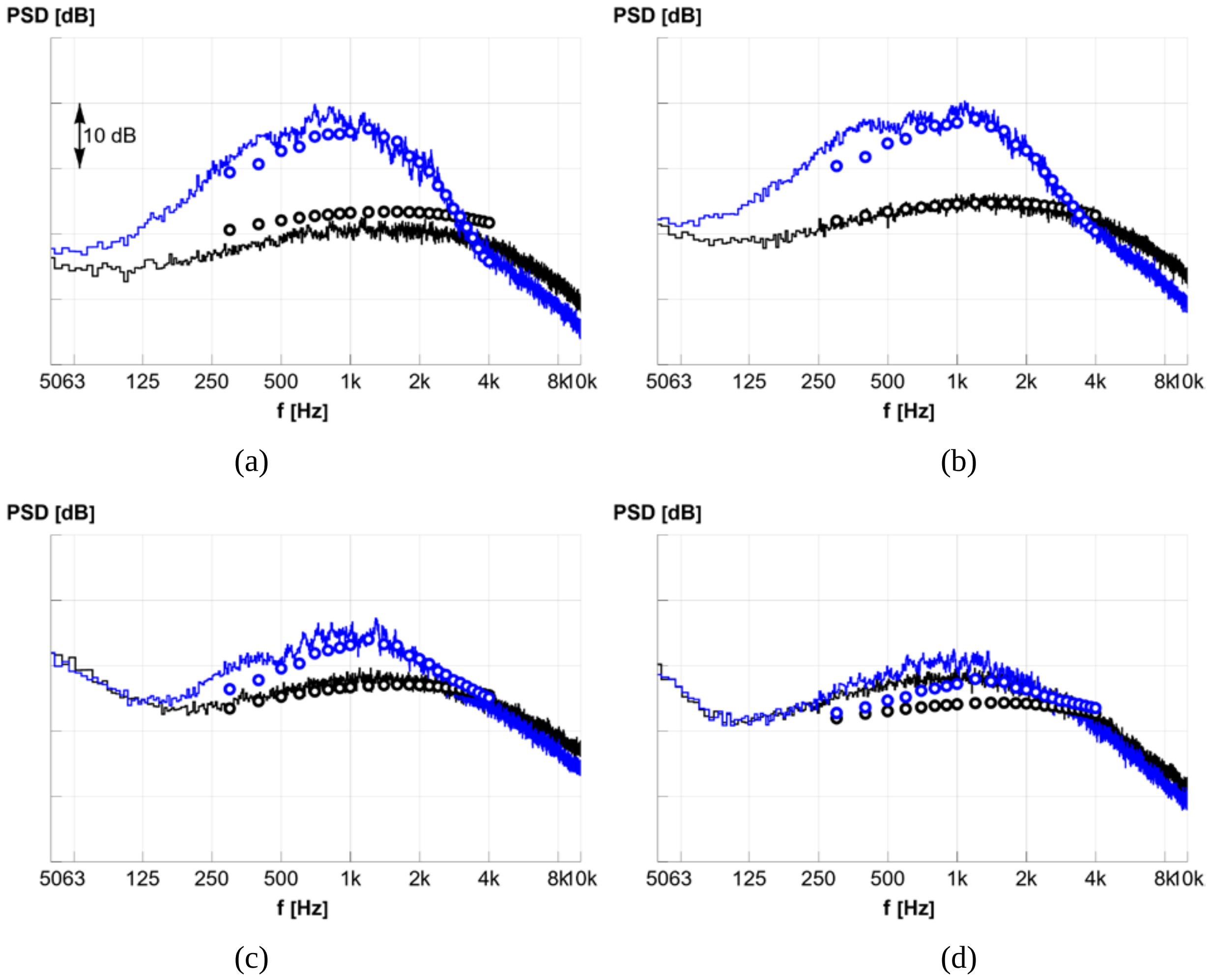

Fig. 7. Spectra of far-field noise for the free jet (black) and installed jet (blue) for different microphone array positions: a) *x/D* = -10; b) *x/D* = 0; c) *x/D* = 18.75; d) *x/D* = 37.5. Solid curves represent the experimental results (microphone 1, Figure 6), markers represent the results obtained by the developed algorithm

It can be noted that the proposed algorithm allows predicting the shape of the spectra and the frequency region of the transition from the dominance of interaction noise to the dominance of jet noise (for example, the intersection of the curves in Figure 7). Some discrepancies can be observed in upstream and downstream directions. This is due to the simplicity of the implemented correlation model neglecting refraction effects and thus overestimating jet noise levels in the upstream direction and underestimating them in the downstream direction.

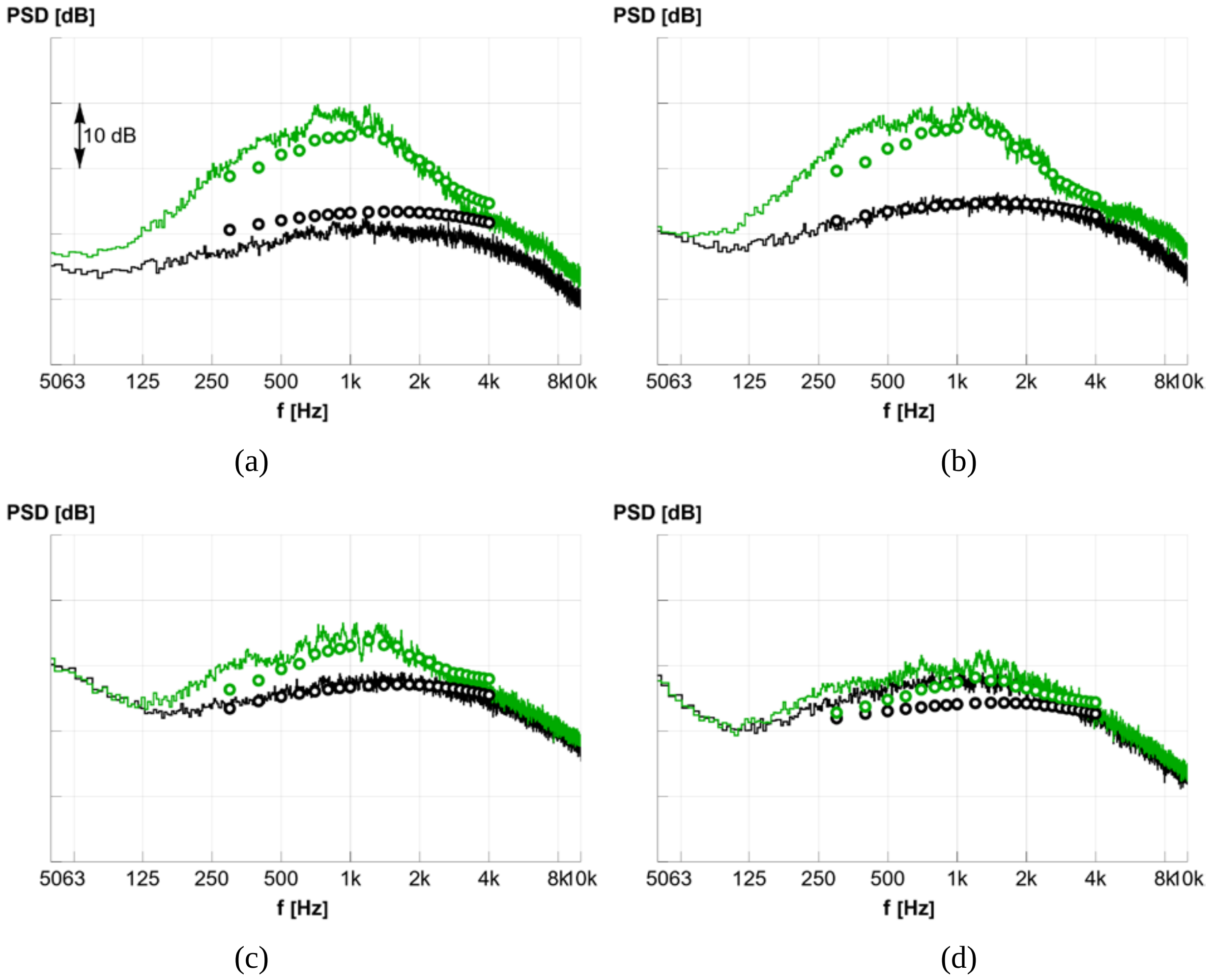

Fig. 8. Spectra of far-field noise for the free jet (black) and installed jet (green) for different microphone array positions: a) $x/D$ = -10; b) $x/D$ = 0; c) $x/D$ = 18.75; d) $x/D$ = 37.5. Solid curves represent the experimental results (microphone 4, Figure 6), markers represent the results obtained by the developed algorithm

A more general picture of the distribution of spectral characteristics on a cylindrical surface is presented in Figure 9 in the form of spectral maps. It can be noted that the algorithm allows to adequately predict the structure of the spectral map of the studied configuration. At the same time, the reduced levels of pulsations in the low-frequency region (300-500 Hz) can be explained by the fact that the scattering model in this preliminary study considers a semi-infinite screen without considering the effect of the leading edge and side edges (e.g., [24]).

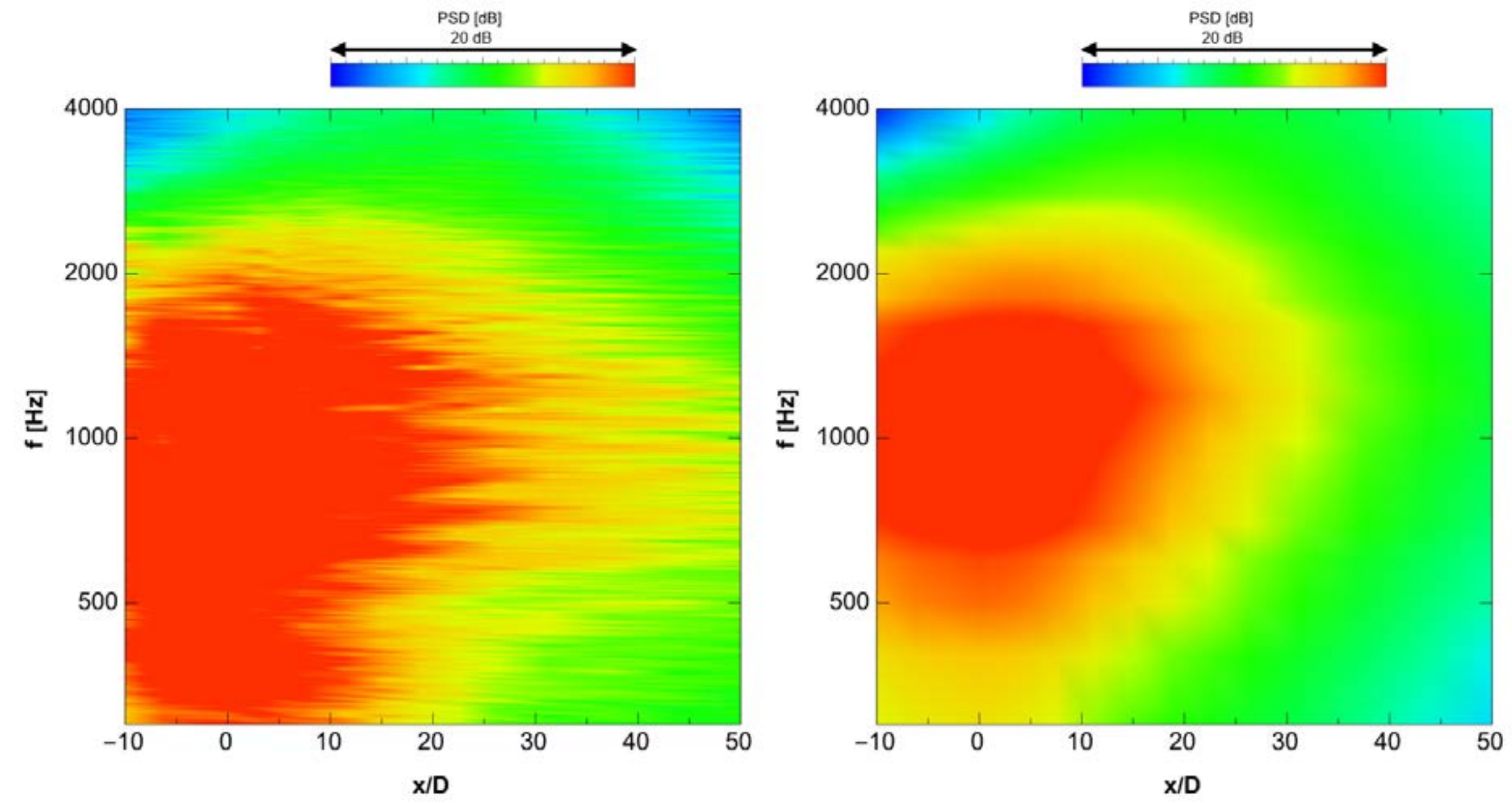

(a) (b)

Fig. 9. Spectral maps of installed jet noise in the shielded region (microphone 1): a) experiment; b) proposed algorithm

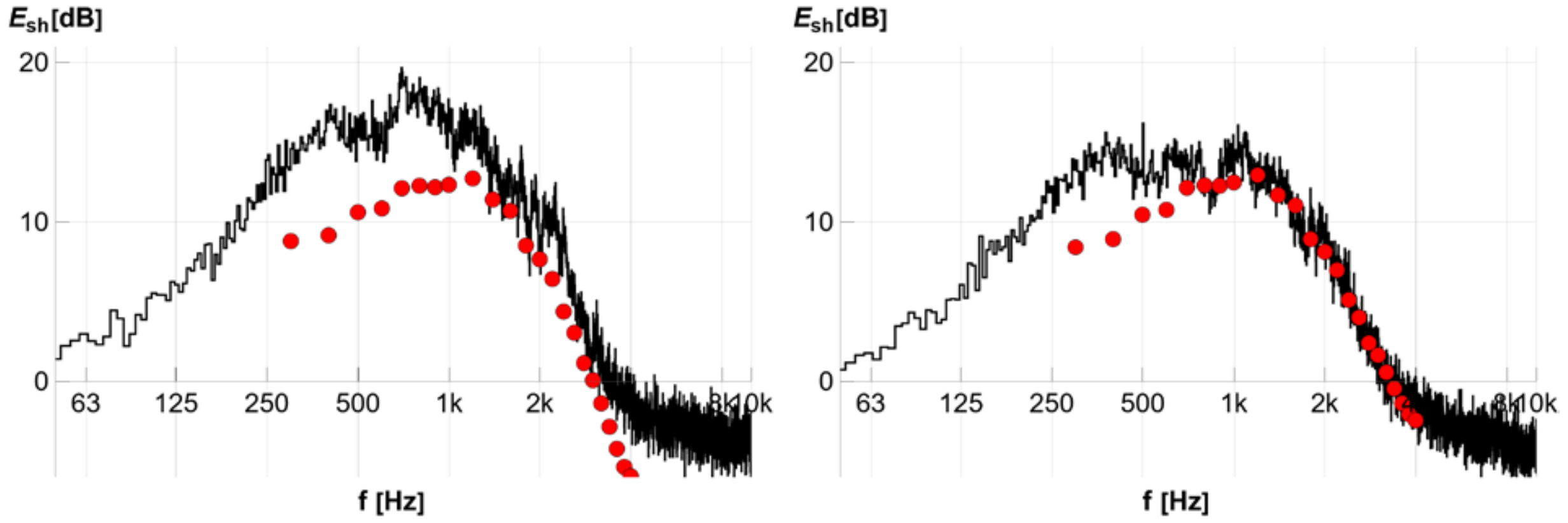

(a) (b)

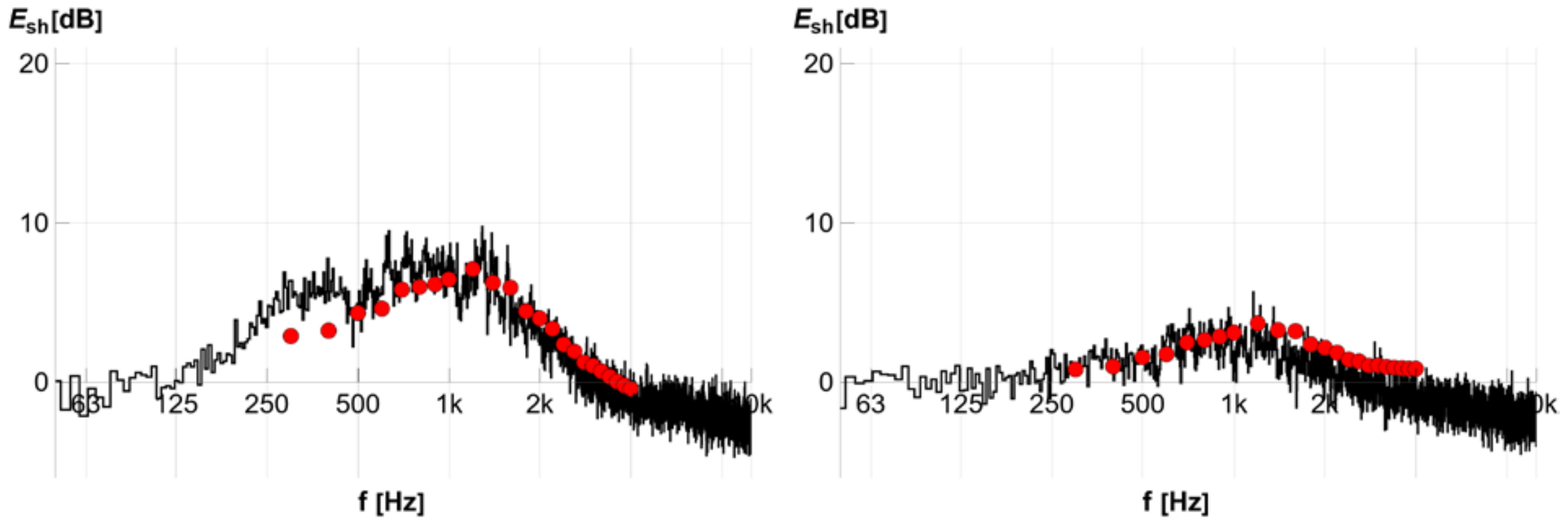

(c) (d)

Fig. 10. Shielding efficiency for different microphone locations: a) *x/D* = -10; b) *x/D* = 0; c) *x/D* = 18.75; d) *x/D* = 37.5. Solid curves represent the experimental results, markers represent the results obtained by the developed algorithm

As indicated above, the developed algorithm allows for an assessment of the shielding efficiency (17). Figure 10 shows a comparison of the shielding efficiency obtained in the experiment and using the algorithm in three characteristic directions of observation.

It can be noted that in the lateral direction the obtained values of shielding efficiency are in very good agreement with the experimental data. The model results for the downstream and upstream directions slightly differ from experimental curves. This situation can be improved by considering a screen of finite dimensions (which is easy to do by GTD approach) and using a more complex isolated jet noise model.

# CONCLUSION

This paper presents the results of a theoretical study of the noise shielding effect of a subsonic turbulent jet located near a rigid screen. The developed theoretical model, based on the methods of the Geometrical Theory of Diffraction, takes into account the effects of scattering of the hydrodynamic fluctuations of the jet near field at the edges of the screen and diffraction of acoustic perturbations generated by small-scale turbulence. Small-scale turbulent noise radiating fluctuations are described within the framework of the correlation theory of jet noise in the form of compact quadrupole sources distributed along the jet axis. Hydrodynamic near-field fluctuations are approximated as a superposition of Kelvin-Helmholtz instability wave packets of different azimuthal numbers. Typical parameters of both types of the fluctuations are adjusted using experimental data on the near and far fields of a free jet issuing from a profiled nozzle with a diameter of 0.04 m at a velocity of 181 m/s. The result jet noise shielding efficiency calculation using the developed model is in good agreement with the experimental data obtained for the jet-plate configuration

From the provided results, it can be concluded that the developed algorithm allows to correctly reproduce the main aeroacoustic effects that occur when the jet stream and the screen are located close to each other: noise amplification in the low-frequency region (interaction noise) and reduction in the high-frequency region (shielding) depending on the relative position of the screen and the nozzle exit section. GTD-based approach also allows considering plane screens of different shapes combining them to simulate more realistic airframe configurations. In the future, a more complex model of the free jet noise, taking into account the effect of refraction, will be implemented in the algorithm to make it applicable for the fast-but-precise calculation the efficiency of jet noise shielding by the surfaces better corresponding to that of real aircraft including innovative concepts with overhead engines.

## FUNDING

The work was carried out within the framework of the RF State Assignment, 28.12.2023, № 020-00005-24-00.